\documentclass[showpacs,aps,prd,nofootinbib,floatfix,amsmath,amssymb]{revtex4}
\usepackage{graphicx}
\usepackage{axodraw}
\begin{document}
\makeatletter
\newbox\slashbox \setbox\slashbox=\hbox{$/$}
\newbox\Slashbox \setbox\Slashbox=\hbox{\large$/$}
\def\pFMslash#1{\setbox\@tempboxa=\hbox{$#1$}
  \@tempdima=0.5\wd\slashbox \advance\@tempdima 0.5\wd\@tempboxa
  \copy\slashbox \kern-\@tempdima \box\@tempboxa}
\def\pFMSlash#1{\setbox\@tempboxa=\hbox{$#1$}
  \@tempdima=0.5\wd\Slashbox \advance\@tempdima 0.5\wd\@tempboxa
  \copy\Slashbox \kern-\@tempdima \box\@tempboxa}
\def\FMslash{\protect\pFMslash}
\def\FMSlash{\protect\pFMSlash}
\def\miss#1{\ifmmode{/\mkern-11mu #1}\else{${/\mkern-11mu #1}$}\fi}
\makeatother

\title{Effective Lagrangian approach to fermion
electric dipole moments induced by a CP--violating $WW\gamma$
vertex}
\author{H. Novales--S\' anchez}
\author{J. J. Toscano}
\email[E-mail:]{jtoscano@fcfm.buap.mx} \affiliation{Facultad de
Ciencias F\'\i sico Matem\'aticas, Benem\'erita Universidad
Aut\'onoma de Puebla, Apartado Postal 1152, Puebla, Pue., M\'
exico.}

\date{\today}

\begin{abstract}
The one--loop contribution of the two CP--violating components of
the $WW\gamma$ vertex, $ \tilde{\kappa}_\gamma W^+_\mu W^-_\nu
\tilde{F}^{\mu \nu}$ and $(\tilde{\lambda}_\gamma / m^2_W)W^+_{\mu
\nu}W^{-\nu}_{\ \ \rho}\tilde{F}^{\rho \mu}$, on the electric
dipole moment (EDM) of fermions is calculated using dimensional
regularization and its impact at low energies reexamined in the
light of the decoupling theorem. The Ward identities satisfied by
these couplings are derived by adopting a $SU_L(2)\times
U_Y(1)$--invariant approach and their implications in radiative
corrections discussed. Previous results on
$\tilde{\kappa}_\gamma$, whose bound is updated to
$|\tilde{\kappa}_\gamma| <5.2\times 10^{-5}$, are reproduced, but
disagreement with those existing for $\tilde{\lambda}_\gamma$ is
found. In particular, the upper bound
$|\tilde{\lambda}_\gamma|<1.9\times10^{-2}$ is found from the
limit on the neutron EDM, which is more than 2 orders of magnitude
less stringent than that of previous results. It is argued that
this difference between the $\tilde{\kappa}_\gamma$ and
$\tilde{\lambda}_\gamma$ bounds is the one that might be expected
in accordance with the decoupling theorem. This argument is
reinforced by analyzing careful the low--energy behavior of the
loop functions. The upper bounds on the $W$ EDM, $|d_W|<6.2\times
10^{-21}\ e\cdot cm$, and the magnetic quadrupole moment,
$|\tilde{Q}_W|<3\times 10^{-36}\ e\cdot cm^2$, are derived. The
EDM of the second and third families of quarks and charged leptons
are estimated. In particular, EDM as large as $ 10^{-20} \ e\cdot
cm$ and $10^{-21} \ e\cdot cm$ are found for the $t$ and $b$
quarks, respectively.
\end{abstract}


\maketitle

\section{Introduction}
\label{Int} Very important information about the origin of CP
violation may be extracted from EDMs of elementary particles. This
elusive electromagnetic property is very interesting, as it
represents a net quantum effect in any renormalizable theory. In
the standard model (SM), the only source of CP violation is the
Cabbibo--Kobayashi--Maskawa (CKM) phase, which however has a
rather marginal impact on flavor--diagonal processes such as the
EDM of elementary particles \cite{DS}. In fact, the EDM of both
fermions and the $W$ gauge boson first arises at the three--loop
level \cite{EDMQW1,EDMQW2}. As far as the magnetic quadrupole
moment of the $W$ boson is concerned, it receives a tiny
contribution at the two--loop level in the SM\cite{MQMW}. Although
they are extremely suppressed in the SM, the EDMs can be very
sensitive to new sources of CP violation, as it was shown recently
for the case of the $W$ boson in a model--independent manner using
the effective Lagrangian technique\cite{TT1,HHPTT}. Indeed, EDMs
can receive large contributions from many SM extensions\cite{BM},
so the scrutiny of these properties may provide relevant
information to our knowledge of CP violation, which still remains
a mystery.

In this paper, we are interested in studying the impact of a
CP--violating $WW\gamma$ vertex on the EDM of charged leptons and
quarks. As it was shown by Marciano and Queijeiro\cite{MQ}, the
CP--odd electromagnetic properties of the $W$ boson can induce
large contributions on the EDM of fermions. Beyond the SM, the
CP--violating $WW\gamma$ vertex can be induced at the one--loop
level by theories that involve both left-- and right--handed
currents with complex phases\cite{TT2,HHPTT}, as it occurs in
left--right symmetric models\cite{LRSM}. Two--loop effects can
arise from Higgs boson couplings to $W$ pairs with undefined CP
structure \cite{TT1}. However, in this work, instead of focusing
on a specific model, we will parametrize this class of effects in
a model--independent manner via the effective Lagrangian
approach\cite{EL}, which is suited to describe those new physics
effects that are quite suppressed or forbidden in the SM. The
phenomenological implications of both the CP--even and the CP--odd
trilinear $WWV$ ($V=\gamma,Z$) couplings have been the subject of
intense study in diverse contexts using the effective Lagrangian
approach \cite{Ha,Wudka}. The static electromagnetic properties of
the $W$ gauge boson can be parametrized by the following effective
Lagrangian:
\begin{equation}
\label{ver}
 {\cal L}_{WW\gamma}=-ie\Big(\Delta \kappa_\gamma F_{\mu
 \nu}W^{-\mu}W^{+\nu}+\frac{\lambda_\gamma}{m^2_W}W^+_{\mu
\nu}W^{-\nu}_{\ \ \rho}F^{\rho \mu}+
 \tilde{\kappa}_\gamma W^+_\mu W^-_\nu
\tilde{F}^{\mu \nu}+\frac{\tilde{\lambda}_\gamma} {m^2_W}W^+_{\mu
\nu}W^{-\nu}_{\ \ \rho}\tilde{F}^{\rho \mu}\Big),
\end{equation}
where $W^\pm_{\mu \nu}=\partial_\mu W^\pm_\nu-\partial_\nu
W^\pm_\mu $ and $\tilde{F}_{\mu \nu}=(1/2)\epsilon_{\mu \nu \alpha
\beta}F^{\alpha \beta}$. The sets of parameters $(\Delta
\kappa_\gamma,\lambda_\gamma)$ and $(\tilde
{\kappa}_\gamma,\tilde{\lambda}_\gamma)$ define the CP--even and
CP--odd static electromagnetic properties of the $W$ boson,
respectively. The magnetic dipole moment ($\mu_W$) and the
electric quadrupole momente ($Q_W$) are defined by \cite{Ha}
\begin{equation}
\mu_W=\frac{e}{2m_W}(2+\Delta \kappa_\gamma+\lambda_\gamma), \ \ \
Q_W=-\frac{e}{m^2_W}(1+\Delta \kappa_\gamma -\lambda_\gamma).
\end{equation}
On the other hand, the electric dipole moment ($d_W$) and the
magnetic quadrupole moment ($\tilde{Q}_W$) are defined by
\cite{Ha}
\begin{equation}
d_W=\frac{e}{2m_W}(\tilde{\kappa}_\gamma+\tilde{\lambda}_\gamma),
\ \ \ \
\tilde{Q}_W=-\frac{e}{m^2_W}(\tilde{\kappa}_\gamma-\tilde{\lambda}_\gamma).
\end{equation}
The dimension--four interactions of the above Lagrangian are
induced after spontaneous symmetry breaking by the following
dimension--six $SU_L(2)\times U_Y(1)$--invariant operators:
\begin{eqnarray}
\label{op1} &&{\cal
O}_{WB}=\frac{\alpha_{WB}}{\Lambda^2}\Big(\Phi^\dag \frac{\sigma^a}{2}\Phi\Big)W^{a\mu \nu}B_{\mu \nu}\\
\label{op11} &&\tilde{{\cal
O}}_{WB}=\frac{\tilde{\alpha}_{WB}}{\Lambda^2}\Big(\Phi^\dag
\frac{\sigma^a}{2}\Phi\Big)W^{a\mu \nu}\tilde{B}_{\mu \nu},
\end{eqnarray}
whereas those interactions of dimension six are generated by the
following $SU_L(2)$--invariants:
\begin{eqnarray}
\label{op2} && {\cal
O}_W=\frac{\alpha_W}{\Lambda^2}\Big(\frac{\epsilon_{abc}}{3!}W^{a\mu}_{\
\ \nu}W^{b\nu}_{\ \ \rho}W^{c\rho}_{\ \ \mu}\Big),
\\
\label{op21}&&\tilde{{\cal
O}}_W=\frac{\tilde{\alpha}_W}{\Lambda^2}\Big(\frac{\epsilon_{abc}}{3!}W^{a\mu}_{\
\ \nu}W^{b\nu}_{\ \ \rho}\tilde{W}^{c\rho}_{\ \ \mu}\Big),
\end{eqnarray}
where $W^a_{\mu \nu}$ and $B_{\mu \nu}$ are the tensor gauge
fields associated with the $SU_L(2)$ and $U_Y(1)$ groups,
respectively. In addition, $\Phi$ is the Higgs doublet, $\Lambda$
is the new physics scale, and the $\tilde{\alpha}_i$ are unknown
coefficients, which can be determined if the underlying theory is
known. As we will see below, the presence of the Higgs doublet in
the ${\cal O}_{WB}$ and $\tilde{{\cal O}}_{WB}$ operators, as well
as its absence in ${\cal O}_W$ and $\tilde{{\cal O}}_W$, has
important physical implications at low energies. In the following,
we will focus on the CP--violating interactions. Introducing the
dimensionless coefficients
$\tilde{\epsilon}_i=(v/\Lambda)^2\tilde{\alpha}_i$, with $v$ the
Fermi scale, it is easy to show that
$\tilde{\kappa}_\gamma=-(c_W/2s_W)\tilde{\epsilon}_{WB}$ and
$\tilde{\lambda}_\gamma=-(e/4s_W)\tilde{\epsilon}_W$, with
$s_W(c_W)$ the sine(cosine) of the weak angle. The impact of the
$\tilde{{\cal O}}_{WB}$ operator on the EDM of fermions, $d_f$,
was studied in Ref.\cite{MQ}. The experimental limits on the EDM
of the electron and neutron were used by the authors to impose a
bound on the $\tilde{\kappa}_\gamma$ parameter \footnote{The
authors of Ref.\cite{MQ} use the symbol $\lambda_W$ to
characterize the $\tilde{F}_{\mu \nu}W^{-\mu}W^{+\nu}$ term, but
more frequently it is used the notation $\tilde{\kappa}_\gamma$,
which we will adopt here.}. It was found that the best bound
arises from the limit on the neutron EDM $d_n$. In this paper,
besides reproducing this calculation using dimensional
regularization and updating the bound on $\tilde{\kappa}_\gamma$
and $d_W$, we will derive an upper bound on the magnetic
quadrupole moment $\tilde{Q}_W$, which, to our knowledge, has not
been presented in the literature.

On the other hand, the contribution of the $\tilde{{\cal O}}_W$
operator to the EDM of fermions has also been studied previously
by several authors \cite{DRS,MANYRS}. Although this operator gives
a finite contribution to $d_f$\footnote{The contribution of the
CP--even ${\cal O}_W$ operator to the magnetic dipole moment of
fermions is also finite \cite{DRS,MANYRS,EW}.}, it has been argued
by the authors of Ref.\cite{MANYRS} that such a contribution is
indeed ambiguous, as it is regularization--scheme dependent. The
authors of Ref.\cite{MANYRS} carried out a comprehensive analysis
by calculating the $\tilde{{\cal O}}_W$ contribution to $d_f$
using several regularization schemes, such as dimensional,
form--factor, Paulli--Villars--regularization, and the Cutoff
method. They show that the result differs from one scheme to
other. In this paper, we reexamine this contribution using the
dimensional regularization scheme. We argue that the result thus
obtained is physically acceptable because it satisfies some low
energy requirements that are inherent to the Appelquist--Carazzone
decoupling theorem \cite{DT}. In particular, we will emphasize the
relative importance of the $\tilde{{\cal O}}_{WB}$ and
$\tilde{{\cal O}}_W$ operators when inserted into a loop to
estimate their impact on a low--energy observable as the EDM of
the electron or the neutron. As we will see below, the
$\tilde{{\cal O}}_{WB}$ operator induces nondecoupling effects,
whereas $\tilde{{\cal O}}_W$ is of decoupling nature. As a
consequence, the constraints obtained from the neutron EDM are
more stringent for $\tilde{{\cal O}}_{WB}$ than for $\tilde{{\cal
O}}_W$, in contradiction with the results of Ref.\cite{MANYRS}
where bounds of the same order of magnitude are found. Below we
will argue on the consistence of our results by analyzing more
closely some peculiarities of these operators in the light of the
decoupling theorem. Although the $\tilde{{\cal O}}_W$ contribution
is insignificant compared with the one of $\tilde{{\cal O}}_{WB}$
for light fermions, it is very important to stress that both
operators can be equally important at high energies. Indeed, we
will see that for the one--loop $\bar{t}t\gamma$ and
$\bar{b}b\gamma$ vertices, the $\tilde{{\cal O}}_W$ contribution
is as large as or larger than the effect of $\tilde{{\cal
O}}_{WB}$. We will see below that as a consequence of the
decoupling nature of $\tilde{{\cal O}}_W$, the bound obtained on
$\tilde{\lambda}_\gamma$ from the neutron EDM is two orders of
magnitude less stringent than that on $\tilde{\kappa}_\gamma$.

Another important goal of this work is to use our bounds on the
$\tilde{\kappa}_\gamma$ and $\tilde{\lambda}_\gamma$ parameters to
predict, besides the CP--odd electromagnetic properties of the $W$
gauge boson, the EDM of the second and third families of charged
leptons and quarks. We believe that the heaviest particles, as the
$W$, $\tau$, $b$, and $t$, could eventually be more sensitive to
new physics effects associated with CP violation. In addition, we
will exploit the $SU_L(2)\times U_Y(1)$ invariance of our
framework to derive limits on the $\tilde{\kappa}_Z$ and
$\tilde{\lambda}_Z$ parameters associated with the CP--odd $WWZ$
vertex.

The paper has been organized as follows. In Sec. \ref{EL}, the
Feynman rule for the CP--odd $WW\gamma$ vertex is presented. We
will focus on the gauge structure of the part coming from the
$\tilde{{\cal O}}_W$ operator. In particular, we will show how
this operator leads to a gauge--independent result even in the
most general case when all particles in the $WW\gamma$ vertex are
off--shell. Sec. \ref{Cal} is devoted to derive the amplitudes for
the on--shell one--loop $\bar{l}l\gamma$ and $\bar{q}q\gamma$
vertices, induced by the CP--odd $WW\gamma$ coupling. In Sec.
\ref{Dis}, the bounds on the $\tilde{\kappa}_\gamma$ and
$\tilde{\lambda}_\gamma$ parameters are derived and used to
predict the EDM of the SM particles. Finally, in Sec. \ref{Con}
the conclusions are presented.

\section{The anomalous CP--violating $WW\gamma$ vertex}
\label{EL}In this section, we present the Feynman rule for the
$WW\gamma$ vertex induced by the effective operators given in Eqs.
(\ref{op11}) and (\ref{op21}). The $\tilde{{\cal O}}_{WB}$ term
can be written in the unitary gauge as follows:
\begin{eqnarray}
\tilde{{\cal
O}}_{WB}&=&-\frac{1}{4}\tilde{\epsilon}_{WB}\tilde{B}_{\mu
\nu}W^{3\mu \nu}+\cdots, \nonumber \\
&=&-ie\Big(\frac{c_W}{2s_W}\tilde{\epsilon}_{WB}\Big)\tilde{F}_{\mu
\nu}W^{-\mu }W^{+\nu}+\cdots,
\end{eqnarray}
where
\begin{eqnarray}
\tilde{B}_{\mu \nu}&=&c_W\tilde{F}_{\mu
\nu}-s_W\tilde{Z}_{\mu \nu}, \\
W^3_{\mu \nu}&=&s_WF_{\mu \nu}+c_WZ_{\mu \nu}+ig(W^-_\mu
W^+_\nu-W^+_\mu W^-_\nu).
\end{eqnarray}
On the other hand, the $\tilde{{\cal O}}_{W}$ term is given by
\begin{eqnarray}
\tilde{{\cal
O}}_{W}&=&\frac{i\tilde{\alpha}_W}{\Lambda^2}\hat{W}^{+\nu
\lambda}\hat{W}^{-\mu}_\nu \tilde{W}^3_{\lambda
\mu} ,\nonumber \\
&=&ie\Big(\frac{e}{4s_W}\tilde{\epsilon}_W\Big)\Big(\frac{1}{m^2_W}\Big)W^{+\nu
\lambda}W^{-\mu}_\nu \tilde{F}_{\lambda \mu}+\cdots,
\end{eqnarray}
where
\begin{equation}
 \hat{W}^+_{\mu \nu}=D_\mu W^+_\nu-D_\nu W^+_\mu+igc_W(W^+_\mu
Z_\nu-W^+_\nu Z_\mu),
\end{equation}
with $D_\mu=\partial_\mu-ieA_\mu$ the electromagnetic covariant
derivative and $\hat{W}^-_{\mu \nu}=(\hat{W}^+_{\mu \nu})^\dag $.

\begin{figure}
\centering
\includegraphics[width=2.0in]{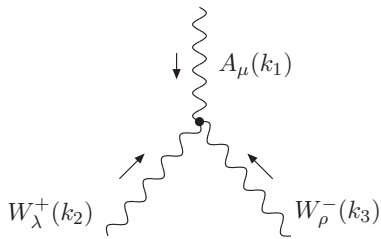}
\caption{\label{FIG1} The trilinear CP--odd $WW\gamma$ vertex.}
\end{figure}

Using the notation shown in Fig.\ref{FIG1}, the vertex function
associated with the $WW\gamma$ coupling can be written as
\begin{equation}
\tilde{\Gamma}^{WW\gamma}_{\lambda \rho
\mu}(k_1,k_2,k_3)=ie\tilde{\kappa}_\gamma \Gamma^{\tilde{\cal
O}_{WB}}_{\lambda \rho \mu}(k_1)+\frac{ie\tilde{\lambda}_\gamma
}{m^2_W}\Gamma^{\tilde{\cal O}_W}_{\lambda \rho \mu}(k_1,k_2,k_3),
\end{equation}
where
\begin{equation}
\Gamma^{\tilde{\cal O}_{WB}}_{\lambda \rho \mu}(k_1)=\epsilon_{\mu
\lambda \rho \eta}k^\eta_1,
\end{equation}
\begin{equation}
\label{vertex}
 \Gamma^{\tilde{\cal O}_W}_{\lambda \rho
\mu}(k_1,k_2,k_3)=(-k_2\cdot k_3\epsilon_{\lambda \rho \mu
\eta}+k_{2\rho}\epsilon_{\lambda \mu \eta
\sigma}k^\sigma_3-k_{3\lambda}\epsilon_{\rho \mu \eta
\sigma}k^\sigma_2)k^\eta_1.
\end{equation}
We now proceed to derive the Ward identities that are satisfied by
these vertex functions. In particular, we will show that as a
consequence of these identities, the $\Gamma^{\tilde{\cal
O}_W}_{\lambda \rho \mu}(k_1,k_2,k_3)$ vertex cannot introduce a
gauge--dependent contribution in any loop amplitude. From
Eq.(\ref{vertex}) it is easy to show that this vertex satisfies
the following simple Ward identities:
\begin{eqnarray}
\label{wi} k^\mu_1\Gamma^{\tilde{\cal O}_W}_{\lambda \rho
\mu}(k_1,k_2,k_3)&=&0, \\
k^\lambda_2\Gamma^{\tilde{\cal O}_W}_{\lambda \rho
\mu}(k_1,k_2,k_3)&=&0, \\
k^\rho_3\Gamma^{\tilde{\cal O}_W}_{\lambda \rho
\mu}(k_1,k_2,k_3)&=&0,
\end{eqnarray}
which arise as a direct consequence of the invariance of
$\tilde{{\cal O}}_W$ under the $SU_L(2)$ group. Since all the
$SU_L(2)\times U_Y(1)$ invariants of dimension higher than four
cannot be affected by the gauge--fixing procedure applied to the
dimension--four theory, any possible gauge dependence necessarily
must arise from the longitudinal components of the gauge field
propagators through the $\xi$ gauge parameter. Gauge--independence
means independence with respect to this parameter. It is clear now
that as a consequence of the above Ward identities, the
$\Gamma^{\tilde{\cal O}_W}_{\lambda \rho \mu}(k_1,k_2,k_3)$
contribution to a multi--loop amplitude is gauge--independent, as
there are no contributions from the longitudinal components of the
propagators and thus it cannot depend on the $\xi$ gauge
parameter. Of course, the complete amplitude maybe
gauge--dependent due to the presence of other gauge couplings.
However, when this anomalous vertex is the only gauge coupling
involved in a given amplitude, as it is the case of the one--loop
electromagnetic properties of a fermion $f$, the corresponding
form factors are manifestly gauge independent. This means that for
all practical purpose, the contribution of this operator to a
given amplitude can be calculated using the Feynman--'tHooft gauge
($\xi=1$). We will see below that, as a consequence of these Ward
identities, the contribution of this operator to the fermion EDM
is not only manifestly gauge--independent, but also free of
ultraviolet divergences. The same considerations apply to the
CP--even counterpart ${\cal O}_W$. These results are also valid
for the $WWZ$ coupling. As far as the $\Gamma^{\tilde{\cal
O}_{WB}}_{\lambda \rho \mu}(k_1)$ vertex is concerned, it also is
subject to satisfy certain Ward identities that arise as a
consequence of the $SU_L(2)\times U_Y(1)$--invariance of the
$\tilde{{\cal O}}_{WB}$ operator. However, these constraints, in
contrast with the ones satisfied by the $\Gamma^{\tilde{\cal
O}_W}_{\lambda \rho \mu}(k_1)$ vertex, are not simple due to the
presence of pseudo Goldstone bosons. These Ward identities are
relations between the $WW\gamma$ and the $G^\pm W^\mp \gamma$
vertices, which are given by
\begin{eqnarray}
k^\mu_1\Gamma^{\tilde{\cal O}_{WB}}_{\lambda \rho
\mu}(k_1)&=&0, \\
k^\lambda_2 \Gamma^{\tilde{\cal O}_{WB}}_{\lambda \rho
\mu}(k_1)&=&m_W\Gamma^{G^+W^-\gamma}_{\mu \rho}(k_1,k_3),\\
k^\rho_3 \Gamma^{\tilde{\cal O}_{WB}}_{\lambda \rho
\mu}(k_1)&=&m_W\Gamma^{G^-W^+\gamma}_{\mu \lambda}(k_1,k_2),
\end{eqnarray}
where
\begin{eqnarray}
\Gamma^{G^+W^-\gamma}_{\mu
\rho}(k_1,k_3)&=&-\frac{1}{m_W}\epsilon_{\mu
\rho \alpha \beta}k^\alpha_3k^\beta_1, \\
\Gamma^{G^-W^+\gamma}_{\mu
\lambda}(k_1,k_2)&=&-\frac{1}{m_W}\epsilon_{\mu \rho \alpha
\beta}k^\alpha_2k^\beta_1.
\end{eqnarray}
These results are also valid for the $WWZ$ vertex. They also apply
to the CP--even counterpart ${\cal O}_{WB}$.

\section{The one--loop induced CP--violating $\bar{f}f\gamma$ vertex}
\label{Cal}
\begin{figure}
\centering
\includegraphics[width=2.0in]{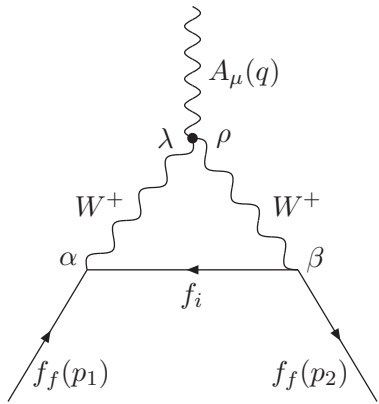}
\caption{\label{FIG3} Contribution of the CP--odd $WW\gamma$
coupling to the on--shell $\bar{f}f\gamma$ vertex.}
\end{figure}

We now turn to calculating the contribution of the
$\tilde{{\cal O}}_{WB}$ and $\tilde{{\cal O}}_W$ operators to the
EDM of a $f$ fermion. The EDM of $f$ is induced at the one--loop
level through the diagram shown in Fig.\ref{FIG3}. It is convenient
to analyze separately the contribution of each operator, as they
possess different features that deserve to be contrasted. To
calculate the loop amplitudes we have chosen the dimensional
regularization scheme, as it is a gauge covariant method which has
probed to be appropriate in theories that are nonrenormalizable in
the Dyson's sense \cite{W}. This framework has been used
successfully in many loop calculations within the context of
effective field theories \cite{MANY}. Although the Feynman
parametrization technique is the adequate method to calculating
on--shell electromagnetic form factors, we will use also the
Passarino--Veltman \cite{PV} covariant decomposition in the case of
the $\tilde{{\cal O}}_W$ contribution, in order to clarify a
disagreement encountered with respect to the results reported in
Ref. \cite{MANYRS}. The Passarino--Veltaman covariant method breaks
down when the photon is on the mass shell, but it can be implemented
with some minor changes~\cite{TT3}.

\subsection{The $\tilde{{\cal O}}_{WB}$ contribution}
We start with the contribution of the $\tilde{{\cal O}}_{WB}$
operator to the on--shell $\bar{f}f\gamma$ vertex. In the
$R_\xi$--gauge, there are contributions coming from the $W$ boson
and its associated pseudo Goldstone boson, but we prefer to use the
unitary gauge in which the contribution is given only through the
diagram shown in Fig.\ref{FIG3}\footnote{The contribution of
$\tilde{{\cal O}}_{WB}$ to reducible diagrams characterized by the
one--loop $Z-\gamma$ mixing vanishes.}. The corresponding amplitude
is given by
\begin{equation}
\Gamma^{\tilde{{\cal
O}}_{WB}}_\mu=\Bigg(\frac{e^3c_W}{4s^3_W}\tilde{\epsilon}_{WB}\Bigg)\epsilon_{\rho
\lambda \mu \sigma}q^\sigma \mu^{4-D}\int
\frac{d^Dk}{(2\pi)^D}\frac{P_R\gamma_\beta
\pFMSlash{k}\gamma_\alpha P^{\alpha \lambda }P^{\beta
\rho}}{\Delta},
\end{equation}
where
\begin{eqnarray}
P^{\alpha \lambda}&=&g^{\alpha \lambda}-\frac{(k+p_1)^\alpha
(k+p_1)^\lambda}{m^2_W}, \\
P^{\beta \rho}&=&g^{\beta
\rho}-\frac{(k+p_2)^\beta(k+p_2)^\rho}{m^2_W},
\end{eqnarray}
\begin{equation}
\Delta=[k^2-m^2_i][(k+p_1)^2-m^2_W][(k+p_2)^2-m^2_W].
\end{equation}
The notation and conventions used in these expressions are shown
in Fig.\ref{FIG3}. It is worth noting that the above amplitude is
divergent, so the integral must be conveniently regularized in
order to introduce a renormalization scheme. The authors of
Ref.\cite{MQ} introduced a cutoff by replacing
$\tilde{\kappa}_\gamma$ with a form factor depending conveniently
on the new physics scale $\Lambda$. Here, as already mentioned, we
will regularize the divergencies using dimensional scheme. As far
as the renormalization scheme is concerned, we will use the
$\bar{MS}$ one with the renormalization scale $\mu=\Lambda$, which
leads to a logarithmic dependence of the form
$\log(\Lambda^2/m^2_W)$. As we will see below, our procedure leads
essentially to the same result given in Ref.\cite{MQ}.

The fermionic EDM form factor $d_f$ is identified with the
coefficient of the Lorentz tensor structure $-i\gamma_5\sigma_{\mu
\nu}q^\nu$. The integrals that arise from the Feynman
parametrization can be expressed in terms of elementary functions.
After some algebra, one obtains
\begin{equation}
\label{tdwb} d^{\tilde{{\cal
O}}_{WB}}_f=-\tilde{\epsilon}_{WB}\Big(\frac{\alpha c_W}{16\pi
s^3_W}\Big)\Big(\frac{e}{2m_W}\Big)\sqrt{x_f}\Big[log\Big(\frac{m^2_W}{\Lambda^2}\Big)+f_{WB}(x_f,x_i)\Big],
\end{equation}
where we have introduced the dimensionless variable
$x_a=m^2_a/m^2_W$. Here, $f_{WB}(x_f,x_i)$ is the loop function,
which is different for leptons or quarks. In the case of charged
leptons, this function is given by
\begin{equation}
\label{lwb}
f_{WB}(x_l)=-\frac{x_l+1}{x_l}+\frac{x^2_l-1}{x^2_l}\log(1-x_l),
\end{equation}
where we have assumed that $x_l<1$. As far as the EDM of quarks is
concerned, the $f_{WB}$ function has a more complicated way, given
by
\begin{eqnarray}
\label{qwb}
f_{WB}(x_q,x_i)&=&-\frac{x_q+x_i+1}{x_q}+\frac{x^2_q+x^2_i-1}{2x^2_q}\log(x_i)+
\nonumber \\
&&\frac{x^3_q-(x_i+1)x^2_q-(x_i+1)^2x_q+(x_i-1)^2(x_i+1)}{2x^2_q\lambda}f(x_q,x_i,\lambda),
\end{eqnarray}
where
\begin{equation}
f(x_q,x_i,\lambda)=\left\{\begin{array}{ll}
\log\Big(\frac{x_i-x_q-\lambda
+1}{x_i-x_q+\lambda+1}\Big), & \textrm{if \ \ $0<x_q<(1+\sqrt{x_i})^2$} \\
\log\Big(\frac{x_i-x_q-\lambda +1}{x_i-x_q+\lambda+1}\Big)-2i\pi ,
& \textrm{if $x_q>(1+\sqrt{x_i})^2$}
\end{array} \right. ,
\end{equation}
with
\begin{equation}
\lambda=\sqrt{x^2_q-2(x_i+1)x_q+(x_i-1)^2}.
\end{equation}
From now on, $m_q$ and $m_i$ will stand for the masses of the
external and internal quarks, respectively.

\subsection{The $\tilde{{\cal O}}_W$ contribution}
We now turn to calculate the contribution of $\tilde{{\cal O}}_W$
to the fermion EDM. In this case, the contribution in the general
$R_\xi$--gauge is given exclusively by the $W$ gauge boson through
the diagram shown in Fig.\ref{FIG3}. Neither pseudo Goldstone
bosons nor ghost fields can contribute, which is linked to the
fact that, as noted previously, there are no contributions from
the longitudinal components of the $W$ propagators due to the
simple Ward identities given in Eq.(\ref{wi}). As a consequence,
the result is manifestly gauge--independent, as any dependence on
the $\xi$ gauge parameter disappears from the amplitude. Also, we
have verified that $\tilde{{\cal O}}_W$ does not contribute to
reducible diagrams characterized by the one--loop $Z-\gamma$
mixing. As already noted by the authors of Ref. \cite{MANYRS}, the
$\tilde{{\cal O}}_W$ operator, in contrast with the $\tilde{{\cal
O}}_{WB}$ one, generates a finite contribution to $d^{\tilde{{\cal
O}}_W}_f$.

As mentioned in the introduction, our result for this operator is
in disagreement with that found in Ref.\cite{MANYRS}. While the
authors of this reference conclude that the loop function
characterizing this contribution is of $O(1)$ in the low--energy
limit (small fermions masses compared with $m_W$), we find that
this function vanishes in this limit. As we will see below, this
leads to a discrepancy of about two orders of magnitude for the
bound on the $\tilde{\lambda}_\gamma$ parameter. It is therefore
important to clarify this point as much as possible. For this
purpose, let us to comment the main steps followed by the authors
of Ref.\cite{MANYRS} in obtaining their result. The starting point
are Eqs.(2.11-2.13), which represent the amplitude for the
contribution of the operator in consideration to the
$\bar{f}f\gamma$ vertex. The next crucial step adopted by the
authors consists in taking the photon momentum equal to zero both
in the numerator and denominator of the integral given by
Eq.(2.11), which leads to the simple expressions given in
Eqs.(3.1,3.2). Next, they use dimensional regularization through
Eqs.(3.5-3.11) to obtain the final result given by Eq.(3.12). This
result comprises the sum of two terms, one which is independent of
the masses involved in the amplitude, and a second term which
vanishes in the low--energy limit. The first term arises from a
careful treatment of the $D\to 4$ limit in dimensional
regularization. We have reproduced all these results. However, we
arrive at a very different result by using only the on--shell
condition, so we think that it is not valid to delete the photon
momentum before carrying out the integration on the momenta space.
We now proceed to show that a different result is obtained if only
the on--shell condition ($q^2=0$ and $q\cdot \epsilon$) is
adopted. Our main result is that the loop function associated with
this operator vanishes in the low--energy limit, in contrast with
the result obtained in Ref.\cite{MANYRS}. To be sure of our
results, we will solve the momentum integral following two
different methods, namely, the Passarino--Veltman~\cite{PV}
covariant decomposition scheme and the Feynman parametrization
technique. After using the Ward identities given in Eq.(\ref{wi}),
the amplitude can be written as follows
\begin{equation}
\Gamma^{\tilde{{\cal
O}}_W}_\mu=-\Big(\frac{e^4}{4!s^3_Wm^2_W}\tilde{\epsilon}_W\Big)\int
\frac{d^Dk}{(2\pi)^D}\frac{P_R\gamma^\rho
\pFMSlash{k}\gamma^\lambda \Gamma^{\tilde{\cal O}_W}_{\lambda \rho
\mu}}{\Delta},
\end{equation}
where $\Gamma^{\tilde{\cal O}_W}_{\lambda \rho \mu}$ represents the
$WW\gamma$ vertex. Once carried out a Lorentz covariant
decomposition, we implement the on--shell condition to obtain:
\begin{eqnarray}
\label{v} d^{\tilde{{\cal
O}}_W}_f&=&-\tilde{\epsilon}_W\Big(\frac{\alpha^{3/2}}{32\sqrt{\pi}s^3_W}\Big)
\Big(\frac{e}{2m_W}\Big)\frac{1}{\sqrt{x_f}}\Bigg[(x_f-1)\Big(B_0(1)-B_0(2)\Big)+2\Big(B_0(3)-B_0(1)\Big)\nonumber
\\
&&+x_i\Big(B_0(1)+B_0(2)-2B_0(3)\Big)+\Big((x_f-x_i)^2-1\Big)m^2_WC_0
\Bigg],
\end{eqnarray}
where $B_0(1)=B_0(m^2_f,m^2_i,m^2_W)$,
$B_0(2)=B_0(0,m^2_W,m^2_W)$, $B_0(3)=B_0(0,m^2_i,m^2_W)$, and
$C_0=C_0(m^2_f,m^2_f,0,m^2_W,m^2_i,m^2_W)$ are Passarino--Veltman
scalar functions. It is important to emphasize that in obtaining
this result, the on--shell condition was implemented only after
calculating the amplitude.

On the other hand, using Feynman parametrization, one obtains
\begin{equation}
\label{p} d^{\tilde{{\cal
O}}_W}_f=-\tilde{\epsilon}_W\Big(\frac{\alpha^{3/2}}{8\sqrt{\pi}s^3_W}\Big)
\Big(\frac{e}{2m_W}\Big)\sqrt{x_f}(I_1+I_2+I_3),
\end{equation}
where the $I_i$ quantities represent parametric integrals, which are
given by
\begin{eqnarray}
I_1&=&\int^1_0dxx(1-x)=\frac{1}{6}, \\
I_2&=&-\int^1_0dx(1-x)(1-3x)\log[(1-x)(1-x_fx)+x_ix], \\
I_3&=&-x_f\int^1_0\frac{dxx^2(1-x)^2}{[(1-x)(1-x_fx)+x_ix]}.
\end{eqnarray}
To clarify our result, let us to analyze more closely these
integrals. The $I_1$ integral, which is independent of the masses,
arises as a residual effect of the $D\to 4$ limit. This apparent
nondecoupling effect that would arise in the low--energy limit is
also found in Ref.\cite{MANYRS}. However, in our case, this effect
is exactly cancelled at low energies by the $I_2$ integral, which
in this limit takes the way:
\begin{equation}
I_2=-\frac{1}{6}+O(x_f,x_i).
\end{equation}
As for the $I_3$ integral, it vanishes in this limit. After solving
the parametric integrals, one obtains
\begin{equation}
\label{tdw} d^{\tilde{{\cal
O}}_W}_f=-\tilde{\epsilon}_W\Big(\frac{\alpha^{3/2}}{16\sqrt{\pi}s^3_W}\Big)
\Big(\frac{e}{2m_W}\Big)\sqrt{x_f}f_W(x_f,x_i),
\end{equation}
where $f_W(x_f,x_i)$ is the loop function. In the case of a charged
lepton, this function is given by
\begin{equation}
\label{lw}
f_W(x_l)=\frac{x_l-2}{x_l}+2\Big(\frac{x_l-1}{x^2_l}\Big)\log(1-x_l).
\end{equation}
On the other hand, the corresponding function for quarks is given
by
\begin{equation}
\label{qw}
f_W(x_q,x_i)=\frac{x_q-2}{x_q}+\frac{x_q+x_i-1}{x^2_q}\log(x_i)+\frac{x^2_q-2x_q+(x_i-1)^2}{x^2_q\lambda}
f(x_q,x_i,\lambda).
\end{equation}
The same result is obtained when the Passarino--Veltman scalar
functions appearing in Eq.(\ref{v}) are expressed in terms of
elementary functions.

In the light of the above results, we can conclude that it is not
valid to delete the photon momentum before carrying out the
integration on the momentum . In the next section, we will argue
that a vanishing loop function in the low--energy limit is the
result that one could expect in accordance with the decoupling
theorem.

\section{Results and discussion}
\label{Dis}We now turn to deriving bounds for the
$\tilde{\epsilon}_{WB}$ and $\tilde{\epsilon}_W$ parameters (or
equivalently, for the $\tilde{\kappa}_\gamma$ and
$\tilde{\lambda}_\gamma$ parameters) using current experimental
limits on the electron and the neutron electric dipole moments. We
will use then these bounds to predict the CP--violating
electromagnetic properties of the $W$ boson and some charged
leptons and quarks.

One important advantage of our approach is that the effective
Lagrangian respects the $SU_L(2)\times U_Y(1)$ symmetry. As a
consequence, the coefficients of the $WW\gamma$ and $WWZ$ vertices
are related at this dimension. The CP--violating part of this
vertex is given by:
\begin{equation}
{\cal L}_{WWZ}=-igc_W\Big(\tilde{\kappa}_Z W^+_\mu W^-_\nu
\tilde{Z}^{\mu \nu}+\frac{\tilde{\lambda}_Z} {m^2_W}W^+_{\mu
\nu}W^{-\nu}_{\ \ \rho}\tilde{Z}^{\rho \mu}\Big),
\end{equation}
where $Z_{\mu \nu}=\partial_\mu Z_\nu-\partial_\nu Z_\mu$. The two
set of parameters characterizing the $WW\gamma$ and $WWZ$
couplings are related by
\begin{eqnarray}
\tilde{\kappa}_Z&=&-\frac{s^2_W}{c^2_W}\tilde{\kappa}_\gamma, \\
\tilde{\lambda}_Z&=&\tilde{\lambda}_\gamma.
\end{eqnarray}
Below, we will constraint both sets of parameters.

The current experimental limits on the electric dipole moments of
the electron and the neutron reported by the particle data book
are \cite{PDG,NB}
\begin{equation}
\label{ede} |d_e|<6.9\times 10^{-28} \ \ \ e\cdot cm,
\end{equation}
\begin{equation}
\label{edn}
 |d_n|< 2.9\times 10^{-26} \ \ \ e\cdot cm.
\end{equation}

\subsection{Decoupling and nondecoupling effects}
Before deriving bounds on the $\tilde{\epsilon}_{WB}$ and
$\tilde{\epsilon}_W$ parameters, let us discuss how radiative
corrections can impact the four Lorentz tensor structures of the
$WW\gamma$ vertex given by Eq.(\ref{ver}). Our objective is to
clarify as much as possible why are so different the bounds that
will be derived below for the $\tilde{\epsilon}_{WB}$ and
$\tilde{\epsilon}_W$ coefficients. First of all, notice that both
$F_{\mu \nu}W^{-\mu}W^{+\nu}$ and $\tilde{F}_{\mu
\nu}W^{-\mu}W^{+\nu}$ terms have a renormalizable structure, as
they are induced by the dimension--four invariants $W^a_{\mu
\nu}W^{a\mu \nu}$ and $\tilde{W}^a_{\mu \nu}W^{a\mu \nu}$.
However, in a perturbative context, only the former of these gauge
invariants remains at the level of the classical action, as the
latter can be written as a surface term. It turns out that, though
renormalizable, the $\tilde{F}_{\mu \nu}W^{-\mu}W^{+\nu}$
interaction arises as a quantum fluctuation and thus it is
naturally suppressed \cite{TT2,LRSM}. The nondecoupling character
of the $\Delta \kappa_\gamma$ and $\tilde{\kappa}_\gamma$ form
factors is well--known from various specific models
\cite{TT2,NDE}. These Lorentz structures can in turn induce
nondecoupling effects when inserted into a loop. In particular,
they can impact significantly low--energy observables, as the EDM
of light fermions. We will show below that this is indeed the
case. In this context, it should be noticed the presence of the
Higgs doublet in the $SU_L(2)\times U_Y(1)$--invariant operators
of Eqs. (\ref{op1},\ref{op11}), which points to a nontrivial link
between the electroweak symmetry breaking scale and these
couplings \cite{Iname}. This connection is also evident in the
nonlinear realization of the effective theory, in which the
analogous of the operators (\ref{op1}) and (\ref{op11}) are:
\begin{eqnarray}
{\cal L}_1&=&a_1gg'Tr\Big(U\frac{\sigma^3}{2}U^\dag
\frac{\sigma^a}{2}\Big)W^{a\mu \nu}B_{\mu \nu},\\
\tilde{{\cal L}}_1&=&a_1gg'Tr\Big(U\frac{\sigma^3}{2}U^\dag
\frac{\sigma^a}{2}\Big)W^{a\mu \nu}\tilde{B}_{\mu \nu}.
\end{eqnarray}
Here, $U=exp\Big(\frac{i\sigma^a \phi^a}{v}\Big)$, with $\phi^a$
the would--be--Goldstone bosons \cite{MJH}. Since in this
model--independent parametrization the new physics is the
responsible for the electroweak symmetry breaking, it is clear
that such a link is beyond the Higgs mechanism. The situation is
quite different for the
 $W^{+\mu}_\nu W^{-\lambda \nu}F_{\mu \lambda}$ and $W^{+\mu}_\nu
W^{-\lambda \nu}\tilde{F}_{\mu \lambda}$ interactions, as they are
nonrenormalizable and thus necessarily arise at one--loop or
higher orders. The decoupling nature of these operators is also
well--known \cite{TT2,NDE}. It is important to notice that the
Lorentz tensor structure of these terms is completely determined
by the $SU_L(2)$ group and that there is no link with the
electroweak symmetry breaking scale, in contrast with the $F_{\mu
\nu}W^{-\mu}W^{+\nu}$ and $\tilde{F}_{\mu \nu}W^{-\mu}W^{+\nu}$
interactions. In this case, it is expected that loop effects of
these operators decouples from low--energy observables. This fact
has already been stressed by some authors \cite{DRujula}. The
reason why these interactions decouple from low--energy
observables stems from the fact that the operators in
Eqs.(\ref{op2},\ref{op21}) respect a global $SU_L(2)$ custodial
symmetry \cite{MANYRS}. We will show below that the loop
contributions of these operators to EDM is of decoupling nature.

We now turn to show the nondecoupling (decoupling) nature of the
$\tilde{F}_{\mu \nu}W^{-\mu}W^{+\nu}$ ($W^{+\mu}_\nu W^{-\lambda
\nu}\tilde{F}_{\mu \lambda}$) contribution to the EDM of light
fermions. We will show that the $f_{WB}$ and $f_W$ loop functions
have a very different behavior for small values of the fermion
masses. We analyze separately the lepton and quark cases. For
fermion masses small compared with the $W$ mass, we can expand the
loop functions given by Eqs.(\ref{lwb},\ref{qwb}) as follows:
\begin{eqnarray}
f_{WB}(x_l)&=&-\frac{1}{2}-\frac{2}{3}x_l+\cdots, \\
f_{WB}(x_q,x_i)&=&-\frac{1}{2}+2x_i-\frac{2}{3}x_q+\cdots
\end{eqnarray}
These results show clearly that the $\tilde{F}_{\mu
\nu}W^{-\mu}W^{+\nu}$ term induces nondecoupling effects. In
practice, this means that a good bound for the
$\tilde{\kappa}_\gamma$ parameter could be derived still from
experimental limits on the EDM of very light fermions, such as the
electron. In contrast with this behavior, as already commented in
the previous section, we can show that $f_W$ is of decoupling
nature:
\begin{eqnarray}
f_W(x_l)&=&-\frac{1}{3}x_l+\cdots ,\\
f_W(x_q,x_i)&=& x_i-\frac{1}{3}x_q+\cdots
\end{eqnarray}
This means that the $\tilde{{\cal O}}_W$ operator only could lead
to significant contributions for heavier fermions. We will show
below that the bound obtained for $\tilde{\lambda}_\gamma$ from
the experimental limit on the EDM of the electron differs in $9$
orders of magnitude with respect to that obtained from the
corresponding limit of the neutron, whereas in the case of the
$\tilde{\kappa}_\gamma$ parameter the analogous bounds differ in
less than 2 orders of magnitude. The high sensitivity of the $f_W$
function to the mass ratios $m_f/m_W$ and $m_i/m_W$ is shown in
Table \ref{TABLE1}. It is interesting to see that $f_W$ and
$f_{WB}$ differ in 4 orders of magnitude for a fermion mass of
about a third of the neutron mass, though they differ in 10 orders
of magnitude for the case of the electron mass. Moreover, notice
that $f_{WB}$ and $f_W$ are of the same order of magnitude for the
third quark family. This means that the $\tilde{{\cal O}}_W$
operator might play an important role in top quark physics. The
very different behavior of the loop functions in the lepton and
quark sectors can be appreciated in Table \ref{TABLE1}. Also, it
should be mentioned that the loop functions develop an imaginary
part in the case of an external quark top. The appearance of an
imaginary (absorptive) part is a consequence of the fact that the
external mass is larger than the sum of the two internal masses:
$m_t>m_W+m_b$.

\begin{table}
\caption{\label{TABLE1} Behavior of the $f_{WB}$ and $f_W$ loop
functions for some specific values of the internal ($m_i$) and
external ($m_f$) fermion masses.}
\begin{ruledtabular}
\begin{tabular}{|l|l|l|l|l|l|}
\hline
$m_f$ & $m_i$ & $x_f$ & $x_i$ & $f_{WB}(x_f,x_i)$  & $f_W(x_f,x_i)$ \\
\hline
 $e$ & $\nu_e$ & $3.9\times 10^{-11} $ & $0$ & $-0. 50$ & $-1. 3\times 10^{-11}$ \\
\hline $\mu$ & $\nu_\mu$ & $1.6 \times 10^{-6}$ & $0$ & $-0.50$ & $-5.2\times 10^{-7}$ \\
\hline $\tau$ & $\nu_\tau$ & $5.0\times 10^{-4}$ & $0$ & $-0.50$ & $-1.6 \times 10^{-4}$\\
\hline $m_n/3$ & $m_n/3$ & $1.5\times 10^{-5}$ & $1. 5\times
10^{-5}$ &
$-0. 49$ & $+1.0\times 10^{-5}$\\
\hline $s$ & $c$ & $2.3 \times 10^{-6}$ & $2.6 \times
10^{-4}$ & $-0. 49$ & $ +2.4\times 10^{-4}$  \\
\hline $c$ & $s$ & $2.6 \times 10^{-4}$ & $2.3 \times
10^{-6}$ & $-0.50$ & $+7.9\times 10^{-5}$  \\
\hline $b$ & $t$ & $2.8 \times 10^{-3}$ & $4.7$ &
$+1.30$ &$+0.6$\\
\hline $t$ & $b$ & $4.7$ & $2.8 \times 10^{-3}$ & $4.7 \times
10^{-2}-3i$ &
$1.02-1.05i$ \\
\end{tabular}
\end{ruledtabular}
\end{table}
\subsection{Bounding the $\tilde{{\cal O}}_{WB}$ operator}
We now turn to deriving a bound for the $\tilde{\kappa}_\gamma$
coefficient using the current experimental limit on the EDM of the
electron and the neutron. In the case of the electron EDM, we can
approximate the $f_{WB}(x_e)$ loop function as follows:
\begin{equation}
f_{WB}(x_e)\approx -\frac{1}{2}-\frac{1}{2}x_e.
\end{equation}
Using this approximation, Eqs. (\ref{tdwb}) and (\ref{ede}) lead
to
\begin{equation}
\Bigg
|\tilde{\epsilon}_{WB}\Bigg(\log\Big(\frac{\Lambda^2}{m^2_W}\Big)+\frac{1}{2}+\frac{1}{2}x_e\Bigg)\Bigg
|<1.3\times 10^{-4}.
\end{equation}
Since in the effective Lagrangian approach one assumes that
$\Lambda \gg m_W$, it is clear that
\begin{equation}log\Big(\frac{\Lambda^2}{m^2_W}\Big)+\frac{1}{2}+\frac{1}{2}x_e>1,
\end{equation}
which allows us to impose the following bound on the  ${\cal
O}_{WB}$ operator
\begin{equation}
|\tilde{\epsilon}_{WB}|<1.6\times 10^{-3},
\end{equation}
which in turn leads to
\begin{equation}
|\tilde{\kappa}_\gamma|<1.5\times 10^{-3}, \ \ \
|\tilde{\kappa}_Z|<4.2\times 10^{-4}.
\end{equation}
In the case of the neutron, as usual, we take $m_u\approx m_d
\approx m_n/3$, with $m_n$ the neutron mass. Also, we assume the
following relation:
\begin{equation}
\label{con} d_n=\frac{4}{3}d_d-\frac{1}{3}d_u.
\end{equation}
Using this connection between the neutron and its constituents,
one obtains for the $\tilde{{\cal O}}_{WB}$ contribution to the
neutron EDM
\begin{equation}
d^{{\tilde{\cal O}}_{WB}}_n=\tilde{\epsilon}_{WB}\Big(\frac{\alpha
c_W}{48\pi
s^3_W}\Big)\Big(\frac{e}{2m_W}\Big)\sqrt{x_n}\Big[\log\Big(\frac{\Lambda^2}{m^2_W}\Big)+f_{WB}(x_n)\Big],
\end{equation}
where
\begin{equation}
f_{WB}(x_n)=2+\frac{9}{x_n}+\frac{81-2x^2_n}{x^2_n}log\Big(\frac{x_n}{9}\Big)
+\frac{4x^2_n+18x_n-81}{2x^2_n\lambda_n}\log\left(\frac{1-\lambda_n}{1+\lambda_n}\right),
\end{equation}
with $\lambda_n=\sqrt{9-4x_n}/3$. Comparing the above theoretical
result with its experimental counterpart given by Eq.(\ref{edn}),
one obtains
\begin{equation}
\Bigg
|\tilde{\epsilon}_{WB}\Bigg(\log\Big(\frac{\Lambda^2}{m^2_W}\Big)+f_{WB}(x_n)\Bigg)\Bigg|<5.
5\times 10^{-5}.
\end{equation}
As in the electron case, it is easy to see that
\begin{equation}
\Bigg |\log\Big(\frac{\Lambda^2}{m^2_W}\Big)+f_{WB}(x_n)\Bigg|>1,
\end{equation}
which allows us to impose the following bound on the $\tilde{{\cal
O}}_{WB}$ operator
\begin{equation}
|\tilde{\epsilon}_{WB}|<5.5 \times 10^{-5},
\end{equation}
which implies
\begin{equation}
|\tilde{\kappa}_\gamma|<5.2\times 10^{-5}, \ \ \
|\tilde{\kappa}_Z|<1.5\times 10^{-5}.
\end{equation}
This bound is almost two orders of magnitude more stringent than
that obtained from the electron EDM. The above results are in
perfect agreement with the ones given in Ref.\cite{MQ}.

\subsection{Bounding the $\tilde{{\cal O}}_W$ operator}
We first explore the possibility of constraining $\tilde{{\cal
O}}_W$ using the experimental limit on the electron EDM. In this
case, a good approximation for the loop function is
$f_W(x_e)\approx -x_e \sim -3\cdot 9\times 10^{-11}$, which in
fact is very small. It leads to a very poor constrain of order of
$10^{7}$. This bound should be compared with the one obtained in
Ref.\cite{MANYRS}, which can be updated to
$|\tilde{\lambda}_\gamma|<7\times 10^{-4}$. This enormous
difference arises because the authors in Ref.\cite{MANYRS} assume
that $f_W\sim O(1)$, instead of $f_W(x_e)\approx -x_e \sim -3.
9\times 10^{-11}$.

We now try to get a more restrictive bound from the experimental
limit on the neutron EDM. Following the same steps given above,
the connection between the EDM of the neutron with its
constituents given in Eq.(\ref{con}) leads to
\begin{equation}
d^{\tilde{{\cal
O}}_W}_n=-\tilde{\epsilon}_W\Big(\frac{\alpha^{3/2}}{48\sqrt{\pi}s^2_W}\Big)
\Big(\frac{e}{2m_W}\Big)\sqrt{x_n}f_W(x_n),
\end{equation}
where
\begin{equation}
f_W(x_n)=2\Big(1-\frac{9}{x_n}\Big)+\frac{9(2x_n-9)}{x^2_n}\log\Big(\frac{x_n}{9}\Big)
+\frac{2x^2_n-36x_n+81}{x^2_n\lambda_n}\log\left(\frac{1-\lambda_n}{1+\lambda_n}\right).
\end{equation}
In this case a more restrictive bound is obtained:
\begin{equation}
\label{n} |\tilde{\epsilon}_W|<0.12,
\end{equation}
which in turn leads to
\begin{equation}
|\tilde{\lambda}_\gamma|=|\tilde{\lambda}_Z|<1.9 \times 10^{-2}.
\end{equation}
In this case the result obtained in Ref.\cite{MANYRS} can be
updated to $|\tilde{\lambda}_\gamma|<6\times 10^{-5}$, which shows
that our constraint is less stringent by more than 2 orders of
magnitude.

From the above results, the high sensitivity of the
$(\tilde{\lambda}_\gamma / m^2_W)W^+_{\mu \nu}W^{-\nu}_{\ \
\rho}\tilde{F}^{\rho \mu}$ interaction to the mass ratio $m_f/m_W$
can be appreciated now.

\subsection{CP--odd electromagnetic properties of fermions and the $W$ gauge boson}
The constraints derived above for the CP--odd $WW\gamma$ vertex
can be used to predict the CP--odd electromagnetic properties of
known particles. In particular, the EDM associated with the
heavier particles are the most interesting, as they could be more
sensitive to new physics effects. Besides the $W$ gauge boson and
the third family of leptons and quarks, we will also include by
completeness the predictions on the members of the second family.
In the case of the $W$ gauge boson, an upper bound for the
magnetic quadrupole moment will also be presented. It should be
emphasized the fact that it is the first time that an upper bound
on $\tilde{Q}_W$ is derived. We will use the constraints derived
from the neutron EDM, as they are most stringent. Since the
$\tilde{{\cal O}}_{WB}$ and $\tilde{{\cal O}}_W$ operators were
bounded one at a time, we will make predictions assuming that the
CP--violating effects cannot arise simultaneously from both
operators. We resume our results in Table \ref{TABLE2}. It should
be noted that while the values for $d_W$ and $\tilde{Q}_W$
constitute true upper bounds, the ones given by the EDM of
fermions are estimations only.

\begin{table}
\caption{\label{TABLE2} Electromagnetic properties of the known
particles induced by a CP--violating $WW\gamma$ vertex. The value
$\Lambda=1000$ GeV is assumed for the contribution of the
$\tilde{{\cal O}}_{WB}$ operator.}
\begin{ruledtabular}
\begin{tabular}{|l|l|l|l|}
\hline
Particle & Electric Dipole Moment& $\tilde{{\cal O}}_{WB}$  & $\tilde{{\cal O}}_W$ \\
\hline $\mu$ & $|d_\mu|$ & $5.7 \times 10^{-26}\ e\cdot cm$ &
$2.0\times 10^{-30}\ e\cdot cm$\\
\hline $\tau$ &  $|d_\tau|$ & $1.0 \times 10^{-24}\ e\cdot cm$ &
$1.1\times 10^{-26}\ e\cdot cm$\\
\hline $c$ &  $|d_c|$ & $7. 1 \times 10^{-25}\ e\cdot cm$ &
$3.8\times 10^{-27}\ e\cdot cm$\\
\hline $s$ & $|d_s|$ & $6.5 \times 10^{-26}\ e\cdot cm$ &
$1.0\times 10^{-27}\ e\cdot cm$\\
\hline $b$ & $|d_b|$ & $2.1 \times 10^{-21}\ e\cdot cm$ &
$9.9\times 10^{-23}\ e\cdot cm$\\
\hline $t$ & $|Re(d_t)|$ & $1.6 \times 10^{-22}\ e\cdot cm$ &
$6.8\times 10^{-21}\ e\cdot cm$\\
\hline $t$ & $|Im(d_t)|$ & $6.0 \times 10^{-23}\ e\cdot cm$ &
$7.0\times 10^{-21}\ e\cdot cm$\\
\hline $W$ & $|d_W|$ & $6.2 \times 10^{-21}\ e\cdot cm$ &
$2.3\times 10^{-18}\ e\cdot cm$\\
\hline $W$ & $|\tilde{Q}_W|$ & $3.0 \times 10^{-36}\ e\cdot cm^2$
&
$1.1\times 10^{-33}\ e\cdot cm^2$\\
\end{tabular}
\end{ruledtabular}
\end{table}

It is worth comparing the limits given in Table \ref{TABLE2} with
some predictions obtained in other contexts. We begin with the
results existing in the literature for the $W$ gauge boson. We
start with the SM predictions for $d_W$ and $\tilde{Q}_W$. As
already mentioned, the lowest order nonzero contribution to $d_W$
arises at the three--loop level, whereas $\tilde{Q}_W$ appears up
to the two--loop order. At the lowest order, $d_W$ has been
estimated to be smaller than about $10^{-29}\ e\cdot cm$
\cite{EDMQW2,PP}. As far as $\tilde{Q}_W$ is concerned, it has
been estimated to be about $-10^{-51} \ e\cdot cm^2$ \cite{MQMW}.
Beyond the SM, almost all studies have focused on $d_W$. Results
several orders of magnitude larger than the SM prediction have
been found. For instance, a value of $10^{-22}\ e\cdot cm$ was
estimated for $d_W$ in left--right symmetric models
\cite{EDMQW2,LRSM} and also in supersymmetric models
\cite{EDMQW2,NC}. Also, a nonzero $d_W$ can arise through
two--loop graphs in multi--Higgs models \cite{SW}. Explicit
calculations carried out within the context of the two--Higgs
doublet model (THDM) show that $d_W\sim 10^{-21}\ e\cdot cm$
\cite{THDM}. A similar value was found within the context of the
so--called 331 models \cite{331}. Recently, the one--loop
contribution of a CP--violating $HWW$ vertex to both $d_W$ and
$\tilde{Q}_W$ was studied in the context of the effective
Lagrangian approach \cite{TT1}. By assuming reasonable values for
the unknown parameters, it was found that $d_W\sim 3-6\times
10^{-21}\ e\cdot cm$ and $\tilde{Q}_W\sim -10^{-36}\ e\cdot cm^2$,
which are 8 and 15 orders of magnitude above the SM contribution.
More recently, the one--loop contribution of the anomalous $tbW$
vertex, which includes both left-- and right--handed complex
components, to $d_W$ and $\tilde{Q}_W$ was calculated
\cite{HHPTT}. By using the most recent bounds on the $tbW$
coupling from $B$ meson physics, it was estimated that $d_W\sim
4\times 10^{-23}-4\times 10^{-22}\ e.cm$ and $\tilde{Q}_W\sim
10^{-38}-10^{-37}\ e.cm^2$. All these predictions for $d_W$ and
$\tilde{Q}_W$ are consistent with the upper bounds given in Table
\ref{TABLE2}.

We now proceed to compare the predictions for the EDM of leptons and
quarks given in Table \ref{TABLE2} with results obtained in some
specific models. As already noted, the values reported for the EDM
of fermions are not upper bounds, as in the case of the $W$ boson,
but only estimates for these quantities, since they are derived by
assuming that CP--violation is induced via a CP--odd $WW\gamma$
vertex. However, it is clear that others sources of CP--violation
could eventually lead to values larger than those presented here.
They are however illustrative of the sensitivity of fermions to
CP--odd effects, so we believe that these results deserve a wider
discussion still in this somewhat restricted scenario. First, we
would like to discuss the prediction existing in the literature for
the $\mu$ and $\tau$ leptons. In the case of the muon, the Particle
Data Group \cite{PDG} reports an experimental limit of about
$d_\mu<10^{-19}\ e\cdot cm$. As far as theoretical predictions are
concerned, the SM prediction is about $10^{-35}\ e\cdot cm$, which
is 16 orders of magnitude below the experimental limit. This means
that precise measurements of the muon EDM might reveal new sources
of CP violation. Although very suppressed in the SM, some of its
extensions predicts values for $d_\mu$ that are several orders of
magnitude larger. For instance, an estimate of $10^{-24}\ e\cdot cm$
for $d_\mu$ was obtained in the THDM \cite{MTHDM}. Similar results
have been found within the context of supersymmetric models
\cite{MSUSY} and in the presence of large neutrino mixing
\cite{MNM}. SUSY model also predict large lepton EDMs if there are
many right-handed neutrinos along with large values of
$tan\beta$~\cite{E}. A wider variety of theoretical perspectives are
studied in \cite{MTP}, where it is found that $d_\mu$ can be as
large as $10^{-22}\ e\cdot cm$. This value is approximately 4 and 8
orders of magnitude above than those induced by the $\tilde{{\cal
O}}_{WB}$ and $\tilde{{\cal O}}_W$ operators, respectively. As far
as the the tau lepton is concerned, the experimental limit is
$-0.22\times 10^{-16}\ e\cdot cm<Re(d_\tau)<0.45\times 10^{-16}\
e\cdot cm$ \cite{PDG}. Since this lepton has a relatively high mass
and a very short lifetime, it is expected that its dynamics is more
sensitive to physics beyond the Fermi scale. Indirect bounds of
order of $|d_\tau|<O(10^{-17})\ e\cdot cm$ have been obtained from
precision LEP data \cite{ESCRIBANO} and naturalness arguments
\cite{PRL}. Some model independent analysis predict possible values
of order $|d_\tau|\sim 10^{-19}\ e\cdot cm$ due to new physics
effects. The possible measurement of $d_\tau$ at low energy
experiments is analyzed in \cite{B}. All these predictions are
consistent with the experimental limit, but are above by at least 7
orders of magnitude with respect to our estimation that arises from
a CP--odd $WW\gamma$ vertex. As far as the EDM of quarks is
concerned, most studies have been focused on the third family. In
the literature, the EDM of the $b$ and $t$ quarks has been
calculated in many variants of multi--Higgs models \cite{MHM}, as it
is expected that more complicated Higgs sectors tend to favor this
class of new physics effects. The dipole moments were estimated to
be of order of $d_b \sim 10^{-23}-10^{-22}\ e\cdot cm$ and $d_t \sim
10^{-21}-10^{-20}\ e\cdot cm$. Very recently, an estimate for $d_t$
of about $10^{-22}\ e\cdot cm$ was obtained from the one--loop
contribution of an anomalous $tbW$ vertex that includes both left--
and right--handed complex components \cite{HHPTT}. It is interesting
to see that in this case the predictions are quite similar to our
estimations derived from the CP--odd $WW\gamma$ vertex. Also, notice
that $\tilde{{\cal O}}_W$ induces the most important contribution.

\section{Conclusions}
\label{Con}The origin of CP violation has remained an unsolved
problem since its discovery several decades ago. Even if the CKM
matrix is the correct mechanism to describe CP violation in $K$ and
$B$ meson systems, this is not necessarily the only source of CP
violation in the nature. Non--zero electric dipole moments of
elementary particles would be a clear evidence of the presence of
new sources of CP violation. In this paper, a source of CP violation
mediated by the $WW\gamma$ vertex has been analyzed using the
effective Lagrangian technique and its implications on the CP--odd
electromagnetic properties of the SM particles studied. Two
dimension--six $SU_L(2)\times U_Y(1)$-- invariant operators,
$\tilde{{\cal O}}_{WB}$ and $\tilde{{\cal O}}_W$, which reproduce
the two independent Lorentz tensor structures, $
\tilde{\kappa}_\gamma W^+_\mu W^-_\nu \tilde{F}^{\mu \nu}$ and
$(\tilde{\lambda}_\gamma / m^2_W)W^+_{\mu \nu}W^{-\nu}_{\ \
\rho}\tilde{F}^{\rho \mu}$, that determine the electric dipole,
$d_W(\tilde{\kappa}_\gamma,\tilde{\lambda}_\gamma)$, and magnetic
quadrupole,
$\tilde{Q}_W(\tilde{\kappa}_\gamma,\tilde{\lambda}_\gamma)$, moments
of the $W$ gauge boson, were introduced. The contribution of this
vertex to the EDM of charged leptons and quarks was calculated. The
main features of these operators were studied in detail. One
interesting peculiarity of the $\tilde{{\cal O}}_W$ operator
consists in the fact that it generates a $WW\gamma$ vertex that
satisfies simple Ward identities. As a direct consequence, the
contribution of this vertex in any multi--loop amplitude is
manifestly gauge--independent. As pointed out by other authors, it
was found that while $\tilde{{\cal O}}_{WB}$ leads to a divergent
amplitude for the fermion EDM, the $\tilde{{\cal O}}_W$ contribution
is free of ultraviolet divergences. The low--energy behavior of
these operators was analyzed in the light of the decoupling theorem.
We emphasized the important fact that while the $\tilde{{\cal
O}}_{WB}$ operator is strongly linked with the electroweak symmetry
breaking (whatever it origin may be), the $\tilde{{\cal O}}_W$ one
has not connection with the electroweak scale. As a consequence, the
former does not decouple at low energies, whereas the latter has a
decoupling nature. Owing to this fact, there is a difference of more
than two orders of magnitude in the respective bounds obtained from
low energy data, in contradiction with previous results given in the
literature where constraints of the same order of magnitude were
derived. The origin of such a disagreement was discussed. At high
energies, the contributions of these operators are equally
important. However, since $\tilde{{\cal O}}_W$ is weakly constrained
by low energy experiments, it might have an important impact on CP
violating observables at high energy collisions. Due to this fact,
$\tilde{{\cal O}}_W$ might be more promising than $\tilde{{\cal
O}}_{WB}$ in searching CP violating effects at high energy
experiments. In order to appreciate these peculiarities, the
behavior of the corresponding loop amplitudes were studied in
detail. The experimental limits on the neutron and electron EDM were
used to get bounds on the $\tilde{\kappa}_\gamma$ and
$\tilde{\lambda}_\gamma$ parameters. It was found that the best
constraints arise from the experimental limit on the neutron EDM,
which leads to $|\tilde{\kappa}_\gamma|<5.2\times 10^{-5}$ and
$|\tilde{\lambda}_\gamma| <1.9\times 10^{-2}$. The former limit
implies the upper bounds $|d_W(5.2\times 10^{-5},0)|<6. 2 \times
10^{-21} \ e\cdot cm$, $|\tilde{Q}_W(5.2\times 10^{-5},0)|<3.
0\times 10^{-36}\ e\cdot cm^2$, whereas the latter leads to
$|d_W(0,1.9\times 10^{-2})|<2.3 \times 10^{-18} \ e\cdot cm$, and
$|\tilde{Q}_W(0,1.9\times 10^{-2})|<1.1\times 10^{-33}\ e\cdot
cm^2$. As far as the limit on $\tilde{\kappa}_\gamma$ and the upper
bound on $d_W$ are concerned, we found agrement with the results
obtained by Marciano and Queijeiro \cite{MQ}. The $SU_L(2)\times
U_Y(1)$ invariance of our approach was exploited to impose
constraints on the $\tilde{\kappa}_Z$ and $\tilde{\lambda}_Z$
parameters associated with the weak coupling $WWZ$. It was found
that $|\tilde{\kappa}_Z|<1.5\times 10^{-5}$ and
$|\tilde{\lambda}_Z|<1.9\times 10^{-2}$. The limits on the
$\tilde{\kappa}_\gamma$ and $\tilde{\lambda}_\gamma$ parameters were
used to estimate the EDM of the muon and tau leptons, as well as the
bottom and top quarks. In the lepton case, we estimated $d_\mu \sim
10^{-26}-10^{-30}\ e\cdot cm$ and $d_\tau \sim 10^{-22}-10^{-26}\
e\cdot cm$, which are 4 and 5 orders of magnitude below than
estimates obtained in other models, respectively. In the case of the
$b$ and $t$ quarks, our estimate is $d_b\sim 10^{-21}-10^{-24}\
e\cdot cm$ and $d_t\sim 10^{-20}-10^{-22}\ e\cdot cm$, which are of
the same order of magnitude than some results found in other
contexts. In general terms, our results indicate that the heavier
fermions, as the $b$ and $t$ quarks, tend to be more sensitive to
new sources of CP violation.

\acknowledgments{We thank G. Tavares--Velasco for his comments.
Financial support from CONACYT and VIEP-BUAP (M\' exico) is also
acknowledged.}


\begin{references}

\bibitem{DS} For a recent review, see M. Pospelov and A. Ritz,
Ann. Phys. (N.Y.) \textbf{318}, 119 (2005).

\bibitem{EDMQW1} M. E. Pospelov and I. B. Khriplovich, Sov. J.
Nucl. Phys. \textbf{53}, 638 (1991) [Yad. Fiz. \textbf{53}, 1030
(1991)]; E. P. Shabalin, Sov. J. Nucl. Phys. \textbf{28}, 75
(1978) [Yad. Fiz. \textbf{28}, 151 (1978)].

\bibitem{EDMQW2} D. Chang, W. Y. Keung, and J. Liu, Nucl. Phys.
\textbf{B355}, 295 (1991).


\bibitem{MQMW} I. B. Khriplovich and M. E. Pospelov, Nucl. Phys.
\textbf{B420}, 505 (1994).

\bibitem{TT1} J. Monta$\tilde{n}$o, F. Ram\'\i rez--Zavaleta, G.
Tavares--Velasco, and J. J. Toscano, Phys. Rev. \textbf{D72},
115009 (2005).

\bibitem{HHPTT} J. Hern\' andez--S\' anchez, C. G. Honorato, F.
Procopio, G. Tavares--Velasco, and J. J. Toscano, Phys. Rev.
\textbf{D75}, 073017 (2007).

\bibitem{BM} S. M. Barr  and W. J. Marciano, in \textit{CP
Violation}, edited by C. Jarlskog (World Scientific, Singapoire,
1989).

\bibitem{MQ} W. J. Marciano and A. Queijeiro, Phys. Rev.
\textbf{D33}, 3449 (1986).

\bibitem{TT2} G. Tavares--Velasco and J. J. Toscano, J. Phys.
\textbf{G30}, 1299 (2004).

\bibitem{LRSM} D. Atwood, C. P. Burgess, C. Hamazaou, B. Irwin,
and J. A. Robinson, Phys. Rev. \textbf{D42}, 3770 (1990).

\bibitem{EL} W. Buchmuller and D. Wyler, Nucl. Phys.
\textbf{B268}, 621 (1986).

\bibitem{Ha} K. Hagiwara, R. D. Peccei, D. Zeppenfeld, and K.
Hikasa, Nucl. Phys. \textbf{B282}, 253 (1987).

\bibitem{Wudka} For a review, see J. Ellison and J. Wudka, Ann.
Rev. Nucl. Part. Sci. \textbf{48}, 33 (1998).

\bibitem{DRS} A. Grau and J. A. Grifols, Phys. Lett.
\textbf{B154}, 283 (1985); J. C. Wallet, Phys. Rev. \textbf{D32},
813 (1985); P. Merry, S. E. Moubarik, M. Perrottet, and F. M.
Renard, Z. Phys. \textbf{C46}, 229 (1990); F. Hoogeveen,
Max--Planck--Institut Report No. MPI-PAE/PTh 25/87, 1987
(unpublished); F. Boudjema, C. P. Burgess, C. Hamzaoui, and J. A.
Robinson, Phys. Rev. \textbf{D43}, 3683 (1991); C. P. Burgess, M.
Frank, and C. Hamzaoui, Z. Phys. \textbf{C70}, 145 (1996).

\bibitem{MANYRS} F. Boudjema, K. Hagiwara, C. Hamzaoui, and K.
Numata, Phys. Rev. \textbf{D43}, 2223 (1991).

\bibitem{EW} C. Arzt, M. B. Einhorn, and J. Wudka, Phys. Rev.
\textbf{D49}, 1370 (1994).

\bibitem{DT} T. Appelquist and J. Carazzone, Phys. Rev.
\textbf{D11}, 2856 (1975).

\bibitem{W} See for instance, J. Wudka, Int. J. Mod. Phys. \textbf{A9}, 2301 (1994).

\bibitem{MANY} See for instance, M. A. P\' erez and J. J. Toscano,
Phys. Lett. \textbf{B289}, 381 (1992); K. Hagiwara, S. Ishihara,
R. Szalapski, and D. Zeppenfeld, Phys. Rev. \textbf{D48}, 2182
(1993); M. A. P\' erez, J. J. Toscano, and J. Wudka, Phys. Rev.
\textbf{D52}, 494 (1995); J. L. D\'\i az--Cruz, J. Hern\'
andez--S\' anchez, and J. J. Toscano, Phys. Lett. \textbf{B512},
339 (2001).

\bibitem{PV} G. Passarino and M. J. G. Veltman, Nucl. Phys.
\textbf{B160}, 151 (1979).

\bibitem{TT3} See for instance, G. Tavares--Velasco and J. J.
Toscano, Phys. Rev. \textbf{D65}, 013005 (2001).

\bibitem{PDG} W. -M. Yao,  \textit{et al.} (Particle Data Group),
J. Phys. \textbf{G33}, 1 (2006).

\bibitem{NB} The most recent bound on the neutron electric
dipole moment is reported in: C. Baker \textit{et al.}, Phys. Rev.
Lett. \textbf{97}, 131801 (2006).

\bibitem{NDE} W. A. Bardeen, R. Gastmans, and B. Lautrup, Nucl.
Phys. \textbf{B46}, 319 (1972); G. Couture and J. N. Ng, Z. Phys.
\textbf{C35}, 65 (1987); G. Couture, J. N. Ng, J. L. Hewett, and T.
G. Rizzo, Phys. Rev. \textbf{D36}, 859 (1987); C. L. Bilachak, R.
Gatsmans, and A. van Proeyen, Nucl. Phys. \textbf{B273}, 46 (1986);
G. Couture, J. N. Ng, J. L. Hewett, and T. G. Rizzo, Phys. Rev.
\textbf{D38}, 860 (1988); A. B. Lahanas and V. C. Sapanos, Phys.
Lett. \textbf{B334}, 378 (1994); T. M. Aliyev, \textit{ibid.},
\textbf{155}, 364 (1985); A. Arhrib, J. L. Kneur, and G. Moultaka,
\textit{ibid.}, \textbf{376}, 127 (1996); N. K. sharma, P. Saxena,
Sardar Singh, A. K. Nagawat, and R. S. Sahu, Phys. Rev.
\textbf{D56}, 4152 (1997); T. G. Rizzo and M. A. Samuel, Phys. Rev.
\textbf{D35}, 403 (1987); A. J. Davies, G. C. Joshi, and R. R.
Volkas, \textit{ibid.}, \textbf{42}, 3226 (1990); F. Larios, J. A.
Leyva, and R. Mart\'\i nez, Phys. Rev. \textbf{D53}, 6686 (1996); G.
Tavares--Velasco and J. J. Toscano, Phys. Rev. \textbf{D69}, 017701
(2004); J. L. Garc\'\i a--Luna, G. Tavares--Velasco, and J. J.
Toscano, Phys. Rev. \textbf{D69}, 093005 (2004); J. Montaño, F.
Ram\'\i rez--Zavaleta, G. Tavares--Velasco, and J. J. Toscano, Phys.
Rev. \textbf{D72}, 055023 (2005); F. Ram\'\i rez--Zavaleta, G.
Tavares--Velasco, and J. J. Toscano, Phys. Rev. \textbf{D75}, 075008
(2007).

\bibitem{Iname} T. Inami, C. S. Lim, B. Takeuchi, and M.
Tanabashi, Phys. Lett. \textbf{B381}, 458 (1996).

\bibitem{MJH} For an introduction to Electroweak Chiral Lagrangians, see Marí\'\i a Jos\' e Herrero,
hep-ph/9601286.

\bibitem{DRujula} A. De R\' ujula, M. B. Gavela, O. Pene, and F. J.
Vegas, Nucl. Phys. \textbf{B357}, 311 (1991); A. De R\' ujula, M.
B. Gavela, P. Hern\' andez, and E. Masso, Nucl. Phys.
\textbf{B384}, 3 (1992).

\bibitem{PP} M. J. Booth, hep-ph/9301293.

\bibitem{NC} I. Vendramin, Nuovo Cimento Soc. Ital. Fis. A
\textbf{105}, 1649 (1992); T. H. West, Phys. Rev. \textbf{D50},
7025 (1994); T. Kadoyoshi and N. Oshimo, Phys. Rev. \textbf{D55},
1481 (1997); N. Oshimo, Nucl. Phys. \textbf{B}, Proc. Suppl.
\textbf{59}, 231 (1997).

\bibitem{SW} S. Weinberg, Phys. Rev. \textbf{D42}, 860 (1990).

\bibitem{THDM} R. Lopez--Mobilia and T. H. West, Phys. Rev.
\textbf{D51}, 6495 (1995); I. Vendramin, Nuovo Cimento Soc. Ital.
Fis. A \textbf{106}, 79 (1993).

\bibitem{331} C. S. Huang and T. J. Li, Phys. Rev. \textbf{D50},
2127 (1994).

\bibitem{MTHDM} V. Barger, A. Das, and C. Kao, Phys. Rev.
\textbf{D55}, 7099 (1997).

\bibitem{MSUSY} T. Ibrahim and P. Nath, Phys. Rev. \textbf{D64},
093002 (2001).

\bibitem{MNM} K. S. Babu, B. Dutta, and R. Mohapatra, Phys. REv.
Lett. \textbf{85}, 5064 (2000).

\bibitem{E} J. R. Ellis and O. Lebedev, Phys. Lett.\textbf{B653}, 411 (2007);
arXiv:0707.4319.

\bibitem{MTP} J. L. Feng, K. T. Matchev, and Y. Shadmi, Nucl.
Phys. \textbf{B613}, 366 (2001).

\bibitem{ESCRIBANO} R. Escribano and E. Mass\' o, Phys. Lett.
\textbf{B301}, 419 (1993); Nucl. Phys. \textbf{B429}, 19 (1994);
Phys. Lett. \textbf{B395}, 369 (1997).

\bibitem{PRL} K. Akama, T. Hattori, and K. Katsuura, Phys. Rev.
Lett. \textbf{88}, 201601 (2002).

\bibitem{B} J. Bernab\' eu, G. A. Gonz\' alez--Sprinberg, and J.
Vidal, Nucl. Phys. \textbf{B763}, 283 (2007).

\bibitem{MHM} S. Weinberg, Phys. Rev. Lett. \textbf{58}, 657
(1976); G. C. Branco and M. N. Robelo, Phys. Lett. \textbf{B160},
117 (1985); J. Liu and L. Wolfenstein, Nucl. Phys. \textbf{B289},
1 (1987); C. H. Albright, J. Smith, and S. H. H. Tye, Phys. Rev.
\textbf{D21}, 711 (1980); A. Soni and R. M. Xu, Phys. Rev. Lett.
\textbf{69}, 33 (1992); N. G. Deshpande and E. Ma, Phys. Rev.
\textbf{D16}, 1583 (1977); Y. Liao and X. Li, Phys. Rev.
\textbf{D60}, 073004 (1999); D. G. Dumm and G. A. Sprinberg, Eur.
Phys. J. \textbf{C11}, 293 (1999); D. A. Demir and M. B. Voloshin,
Phys. Rev. \textbf{D63}, 115011 (2001); E. O. Iltan, J. Phys.
\textbf{G27}, 1723 (2001); Phys. Rev. \textbf{D65}, 073013 (2002).

\end{references}
\end{document}